\documentclass[authoryear,preprint,review,12pt]{elsarticle}

\usepackage{amsthm}
\usepackage{amsmath,amssymb,amsfonts}
\usepackage{graphicx,psfrag,epsf}
\usepackage{enumerate}
\usepackage{mathtools}
\usepackage[authoryear]{natbib}
\usepackage{algorithm}
\usepackage{algorithmic}
\usepackage{booktabs}
\usepackage{multirow}
\usepackage{url} 
\usepackage[colorlinks,
linkcolor=red,
anchorcolor=green,
citecolor=blue
]{hyperref}
\usepackage[font=normalsize]{caption}
\usepackage{makecell}
\usepackage{rotating}
\usepackage{subcaption}

\theoremstyle{plain}
\newtheorem{theorem}{Theorem}

\newtheorem{proposition}[theorem]{Proposition}

\theoremstyle{remark}
\newtheorem{definition}{Definition}

\newtheorem{remark}{Remark}

\def\spacingset#1{\renewcommand{\baselinestretch}%
{#1}\small\normalsize} \spacingset{1}

\usepackage{lineno}

\journal{arXiv}

\begin{document}

\begin{frontmatter}
	

\title{Parsimonious Generative Machine Learning for Non-Gaussian Tail Modeling}

\author[label1]{Xing Yan\corref{cor2}}
\ead{yanxing128@gmail.com}
\author[label2]{Yue Zhao\corref{cor2}}
\ead{zhaoyue123@ruc.edu.cn}
\cortext[cor2]{Co-first authors.}
\author[label3]{Qi Wu}
\ead{qi.wu@cityu.edu.hk}
\author[label4]{Wenxuan Ma\corref{cor1}}
\ead{wenxuanma@btbu.edu.cn}
\cortext[cor1]{Corresponding author.}
\address[label1]{Beijing Institute of Mathematical Sciences and Applications (BIMSA), Beijing, China}
\address[label2]{Institute of Statistics and Big Data, Renmin University of China, Beijing, China}
\address[label3]{Department of Data Science, City University of Hong Kong, Hong Kong SAR}
\address[label4]{School of Mathematics and Statistics, Beijing Technology and Business University, Beijing, China}

\begin{abstract}
The presence of non-Gaussian tails is a prevalent characteristic in many financial modeling scenarios, necessitating the use of complex non-Gaussian distributions such as the generalized beta of the second kind (GB2) and the skewed generalized $t$ (SGT). The approach we propose for modeling heavy-tailed data differs significantly from traditional methods. We utilize generative machine learning, which offers an entirely different paradigm for modeling distributions. A parsimonious nonlinear transformation is applied to a simple base random variable such as Gaussian. The parameters can be estimated effectively, and the theoretical heavy-tail properties are derived. Robust performance is observed with this approach when compared to traditional distributions. More importantly, this method is broadly useful for machine learning due to its mathematical elegance and numerical convenience.
\end{abstract}



\begin{keyword}
Non-Gaussian Tails \sep
Generative Machine Learning \sep
Quantile Regression \sep
GB2 \sep
SGT
\end{keyword}

\end{frontmatter}

\newpage

\section{Introduction} \label{sec:introduction}


It is well-documented that non-Gaussian tails are prevalent in financial markets \citep{rachev2003handbook, jondeau2007financial}. Furthermore, these tails are typically asymmetric between the left and right sides \citep{cont2001empirical, mcneil2015quantitative}. In response to the limitations of the Gaussian assumption, several advanced distribution families have been developed and widely used, such as the four-parameter generalized beta of the second kind (GB2) \citep{mcdonald2008some}, the five-parameter skewed generalized $t$ (SGT) \citep{theodossiou1998financial}, and various special cases of them. 

Without exception, these distributions are defined by relatively complex density functions, which limit their practical flexibility in real-world applications. While their analytical tractability is an advantage, they come with significant drawbacks, including challenges in sampling, difficulty in density computation, and complications in calculating expectations. In the context of machine learning, particularly deep learning \citep{lecun2015deep}, the need to take derivatives of the loss function with respect to trainable parameters is common, yet the derivatives of these distributions with respect to their parameters are often difficult to obtain, further restricting their applicability in machine learning.


Inspired by these considerations and motivated by the recent successes of machine learning, we propose to address the problem using a prevalent and promising approach known as generative machine learning. Specifically, we introduce a parsimonious parametric model capable of describing rich classes of distributions by transforming a simple base random variable, such as the Gaussian-distributed. The design of this transformation takes into account the presence of asymmetric non-Gaussian tails. Although the model lacks a closed-form density, its parameters can be effectively estimated using quantile regression. We apply this approach to model asset returns in financial markets, demonstrating the merits of the proposed model through extensive numerical results presented in this paper. Additionally, this model is well-suited for machine learning tasks, such as conditional distribution prediction, as the derivatives with respect to the distribution parameters can be readily obtained.

This paper is organized as follows. In Section \ref{sec:gml}, we review the general concept of generative machine learning for distribution modeling. Section \ref{sec:generalization} provides a detailed introduction to the proposed method. In Section \ref{sec:parameter}, we outline the procedure for parameter estimation. Section \ref{sec:applications} presents the numerical results of the proposed model, along with comparisons to traditional methods, highlighting its effectiveness. Finally, Section \ref{sec:conclusion} concludes the paper.

\section{Generative Machine Learning}
\label{sec:gml}

Recent advancements in generative machine learning (GML) models have introduced a fundamentally different paradigm for distribution modeling. In this framework, the focus shifts away from directly modeling the density function or distributional function. Instead, a novel methodology based on simple variable transformation is employed. Specifically, given a simple random variable or vector $Z$ that follows a well-known simple distribution (e.g., Gaussian), the target variable(s) $Y$ of interest (e.g., the observations in a dataset) is assumed to be generated according to a transformation:
\begin{equation}\label{GML basic}
	Y=F_\Theta (Z),
\end{equation}
where $F_\Theta$ is usually a deep neural network with trainable parameters $\Theta$, which must be learned such that $Y$ is distributed in a manner closely resembling the observed data. Different generative models employ various choices for $F_\Theta$ and adopt distinct learning strategies for $\Theta$. Among the most commonly used models are Variational Autoencoders (VAEs) \citep{kingma2013auto, kingma2019introduction}, Generative Adversarial Networks (GANs) \citep{GAN, arjovsky2017wasserstein}, Normalizing Flows (NFs) \citep{kobyzev2020normalizing, NormalizingFlow}, diffusion models \citep{rombach2022high, croitoru2023diffusion, yang2023diffusion}, and their numerous variants.


A key advantage of generative models over traditional density-based distribution modeling approaches is their convenience for sampling. Despite the absence of a closed-form density function, it is straightforward to sample from Equation \eqref{GML basic} by generating a realization of $Z$ and passing it through $F_\Theta$. This feature is particularly valuable in the financial domain, where Monte Carlo simulations are frequently required, especially in derivatives markets. In contrast, density-based sampling usually introduces challenges.

Besides, the density-based approach faces challenges in the optimization process, particularly in gradient-based optimization, due to the complexity of many densities with asymmetric non-Gaussian tails. These densities often have intricate mathematical forms, making it difficult to compute derivatives with respect to distribution parameters. When estimating conditional distributions or when calculating and optimizing expectations (or moments, as in portfolio optimization), the limitations of these complicated densities become evident.

The generative models discussed above have achieved significant success in modeling and generating data across domains such as image analysis, time series, and so on. These models typically employ a deep neural network as $F_\Theta$, as the data in these cases is often high-dimensional and exhibits a low-dimensional manifold structure. However, this assumption may not hold in the context of this paper. In financial markets, a highly risk-sensitive environment, modeling a distribution requires a nuanced description of how it deviates from the Gaussian, particularly with respect to tail behavior and asymmetry. \cite{yan2018parsimonious} proposed a non-Gaussian distribution in the form of a quantile function for Value-at-Risk (VaR) prediction in financial markets. This approach is equivalent to choosing a standard Gaussian-distributed $Z$ in Equation (\ref{GML basic}) and carefully designing $F_\Theta$ to induce asymmetric left and right non-Gaussian tails in $Y$. We adopt this idea in this paper for modeling financial returns, as it offers unique advantages, which will be demonstrated in the subsequent sections.

\subsection{An Existing Parsimonious GML Model}

Originally proposed by \cite{yan2018parsimonious} and further refined by \cite{yan2019cross}, researchers introduced a novel parametric quantile function to represent a univariate distribution for modeling asset returns. This function is a monotonically increasing nonlinear transformation of the standard Gaussian quantile function $\Phi^{-1}(\alpha), ~\alpha \in(0,1)$:
\begin{equation}\label{htqf}
	Q(\alpha \mid \mu, \sigma, u, v)=\mu+\sigma \Phi^{-1}(\alpha)\left(\frac{u^{\Phi^{-1}(\alpha)}}{A}+\frac{v^{-\Phi^{-1}(\alpha)}}{A}+1\right),
\end{equation}
where $\mu$ and $\sigma$ represent the location and scale, respectively, while $A$ (set to 4) is a hyperparameter. The resulting distribution associated with the quantile function $Q(\alpha \mid \mu, \sigma, u, v)$ exhibits heavier non-Gaussian tails, with the right and left tail weights controlled by the parameters $u \geq 1$ and $v \geq 1$, respectively. Actually, a reversed S-shape can be observed in the Q-Q plot of $Q(\alpha \mid \mu, \sigma, u, v)$ against $\Phi^{-1}(\alpha)$. Specifically, the parameter $u$ governs the upper tail of the inverted S-shape, corresponding to the right tail of the distribution, while $v$ controls the lower tail, representing the left tail. As these parameters increase, the tails become heavier. Setting $u$ or $v$ to 1 results in a standard Gaussian tail.


Due to the monotonic nature of the transformation on $\Phi^{-1}(\alpha)$, the inverse of $Q(\alpha \mid \mu, \sigma, u, v)$ is uniquely determined and represents a valid distribution. Furthermore, because of this monotonic property, the methodology is equivalent to applying the same transformation, 
\begin{equation}\label{htqf.generative}
	Y=\mu+\sigma Z\left(\frac{u^{Z}}{A}+\frac{v^{-Z}}{A}+1\right), 
\end{equation}
to the standard Gaussian variable $Z \sim \mathcal{N}(0, 1)$. The random variable $Y$ is exactly characterized by the quantile function $Q(\alpha \mid \mu, \sigma, u, v)$. Moreover, parameter estimation based on observed data can be performed using quantile regression, as described by Equation \eqref{htqf}.

\section{Generalization and Tail Heaviness}
\label{sec:generalization}

In this section, we extend the previous model to allow for more flexible asymmetric non-Gaussian tails. The model discussed earlier is based on the fact that a strictly increasing transformation of a random variable is equivalent to transforming its quantile function. Our aim here is to define transformations that can easily accommodate flexible tail properties and asymmetry characteristics through intuitive parameters. To begin, we introduce our definition of relative tail heaviness.


Given a continuous cumulative distribution function $F$, we denote its survival function as $\bar{F}(x) = 1 - F(x)$. Without loss of generality, we assume that $F$ is strictly increasing over an infinite interval $(x_r,+\infty)$, ensuring that the quantile function $F^{-1}$ exists and is also continuous and strictly increasing on $(\alpha_r,1)$. In this way, we can focus solely on the right-tail behavior. Additionally, we use the notation $f(x) \sim g(x)$ to indicate that $\lim_{x \to +\infty} \frac{f(x)}{g(x)}$ is a positive constant.

\begin{definition}
	We say that two distributions have the same right-tail heaviness if their survival functions satisfy $\bar{F_1}(x) \sim \bar{F_2}(x)$ as $x \to +\infty$. Conversely, we say that one distribution $F_2$ has a heavier right tail than another distribution $F_1$ if their survival functions satisfy:
	\begin{equation}
		\lim_{x\to +\infty} \frac{\bar{F_2}(x)}{\bar{F_1}(\frac{x-\mu}{\sigma})} = + \infty, \quad \forall \mu\in\mathbb{R} \text{ and } \sigma >0.
	\end{equation}
\end{definition}

\begin{definition}\label{def.tail_heaviness}
	We also say that two distributions have the same right-tail heaviness if there exist $\mu \in \mathbb{R}$ and $\sigma > 0$ such that $\bar{F_1}((x-\mu)/\sigma)\sim \bar{F_2}(x)$ as $x\to +\infty$.
\end{definition}

The first definition suggests that tail heaviness implies that $\bar{F_2}$ approaches zero more slowly than any location-scale transformation of $\bar{F_1}$, which aligns with the location-scale transformation of the corresponding random variable. The second definition is an extension, with intuitive examples including the fact that all Gaussian distributions share the same tail heaviness, all exponential distributions share the same tail heaviness, and all distributions with a fixed tail index exhibit the same tail heaviness. It is straightforward to show that this definition of equivalent tail heaviness forms an equivalence relation algebraically. The proof is simple and thus omitted.



\subsection{A Generalized Form of Transformation}\label{quantile-transformation}

The approach we adopt to define a new distribution in this work differs significantly from traditional methods. In brief, we construct a novel distribution with arbitrary tail heaviness by specifying its quantile function through a transformation of a known base quantile function, or equivalently, via a random variable transformation. We begin with the general concept and present our method in the most general form, followed by specific examples to illustrate its flexibility.

We propose the following generalized form of transformation.
\begin{definition}\label{def:def1}
	We define a function $y = f(x), ~ x \in \mathbb{R}$, which is described as follows, to curve the straight line $y = \mu + \sigma x$ in the $(x, y)$-coordinate plane:
	\begin{equation}
		f(x) = \mu+\sigma x \left(g_1(x)+g_2(x)+1\right),
		\label{eqn:eqn_f}
	\end{equation}
	where $\sigma>0$, and $g_1(x)$ and $g_2(x)$ satisfy the following conditions:
	\begin{enumerate}
		\item both functions are differentiable at $x\ne 0$ and continuous at $x\in\mathbb{R}$,
		\item $g_1(x)$ is monotonically non-decreasing, while $g_2(x)$ is monotonically non-increasing,
		\item $\lim_{x\to-\infty}g_1(x)=0$ and $\lim_{x\to+\infty}g_2(x)=0$.
	\end{enumerate}
\end{definition}

In this definition, it is evident that $g_i(x)\ge0,~\forall x\in\mathbb{R},~i=1,2$. Therefore, $f(x)$ approaches $-\infty$ as $x\rightarrow -\infty$ and $+\infty$ as $x\rightarrow +\infty$. Furthermore, $f(x)$ is differentiable over $(-\infty,+\infty)$. At $x=0$, we have $f'(0)=\lim_{x\to 0}\sigma(g_1(x)+g_2(x)+1)=\sigma(g_1(0)+g_2(0)+1)$. Next, we prove the monotonicity of $f(x)$. If $f(x)$ is strictly increasing, then $f^{-1}(x)$ exists uniquely on $(-\infty,+\infty)$ and is differentiable.
\begin{proposition}\label{prop:prop1}
	If $g_1(x)$ and $g_2(x)$ satisfy the three conditions in Definition \ref{def:def1}, along with an additional condition: 
	\begin{equation}
		g_i(x)+xg_i'(x)>-\frac{1}{2},~\forall x\ne 0,~i=1,2,
	\end{equation}
	then $f(x)$ in Equation (\ref{eqn:eqn_f}) is strictly increasing.
\end{proposition}
\begin{proof}
	It can be demonstrated that 
	\begin{equation}
		f'(x)=\sigma\left(g_1(x)+g_2(x)+1\right)+\sigma x \left(g_1'(x)+g_2'(x)\right)>0,
	\end{equation}
	at $x\ne 0$, and $f'(0)=\sigma(g_1(0)+g_2(0)+1)>0$. This completes the proof.
\end{proof}

This additional condition can easily be satisfied, allowing for considerable flexibility in the choice of $g_i(x)$. For instance, for $g_1(x)$, the inequality $g_1(x) + xg_1'(x) \ge 0 > -\tfrac{1}{2}$ always holds when $x > 0$. For $x < 0$, one can select a function $g_1(x)$ with a sufficiently small derivative near zero and satisfying $\lim_{x \to -\infty} xg_1'(x) = 0$. A simple choice is $g_1(x) = \mathbb{I}_{\{x \ge 0\}}x^u$, where $u > 1$. The function $g_2(x)$ can be defined symmetrically, possibly with a different parameter $u$. Additional examples of $g_1(x)$ and $g_2(x)$ will be provided in a later subsection. 

Intuitively, the function $y = f(x)$ bends the straight line $y = \mu + \sigma x$ into an inverted S-shape with fattened tails, where $g_1(x)$ and $g_2(x)$ operate on the upper and lower sides, respectively. Next, we apply this transformation to a base random variable, or equivalently, to its quantile function.

\begin{proposition}
	Let $F(x)$ denote the cumulative distribution function of a random variable $X$, and let $F^{-1}(\alpha)$, $\alpha \in (0,1)$, denote its quantile function. If $f(x)$ satisfies the four conditions stated in Proposition \ref{prop:prop1}, then the transformed function $f(F^{-1}(\alpha))$ is the quantile function of the new random variable $Y = f(X)$.
\end{proposition}
\begin{proof}
The result follows directly from the three properties of $f(x)$: it is strictly increasing, differentiable, and approaches $-\infty$ and $+\infty$ as $x$ tends to $-\infty$ and $+\infty$, respectively.
\end{proof}

This yields a convenient simulation property, which constitutes a major advantage over traditional heavy-tail modeling approaches. To generate samples from the distribution of $Y$, one simply draws samples from the base distribution of $X$ and applies the transformation $f$. This computational simplicity is particularly valuable in financial problems such as derivative pricing and risk management. Moreover, parameter estimation—whether unconditional or conditional—can be effectively performed using quantile regression, as quantiles of $Y$ are readily computable given the simple base distribution of $X$.

If one plots $f(F^{-1}(\alpha))$ against $F^{-1}(\alpha)$, the resulting curve corresponds precisely to a Q–Q plot, which is commonly used to compare the tail behaviors of two distributions. The transformation $f(x)$ fully determines the shape of this Q–Q plot. Consequently, we have constructed a new distribution by specifying its quantile function through the prescribed shape of its Q–Q plot relative to a simple base distribution.

\subsection{Some Examples}
\label{sec:examples}

Here, the distribution $F(x)$ is referred to as the base distribution. Since our primary interest lies in the tail behavior of the new distribution, we require that the base distribution possess a nonzero density on at least one of the two infinite sides (left or right). The functions $g_1(x)$ and $g_2(x)$ then serve to control the degree of tail heaviness relative to the base distribution: the faster $g_1(x)$ and $g_2(x)$ go to infinity, the heavier the resulting tails become. Throughout the remainder of this paper, we always assume that $f(x)$ satisfies the four conditions specified in Proposition \ref{prop:prop1}.

\begin{figure}[t]
	\begin{center}
		\begin{subfigure}[t]{.25\linewidth}
			\centerline{\includegraphics[width=\linewidth]{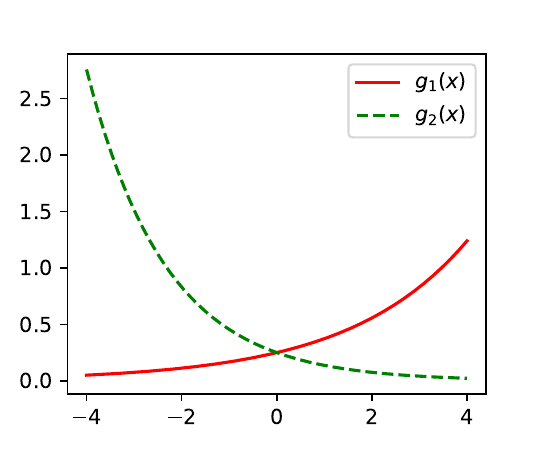}}
			\caption{{\footnotesize $g_1(x)$ and $g_2(x)$}}
		\end{subfigure}\hspace{1em}
		\begin{subfigure}[t]{.25\linewidth}
			\centerline{\includegraphics[width=\linewidth]{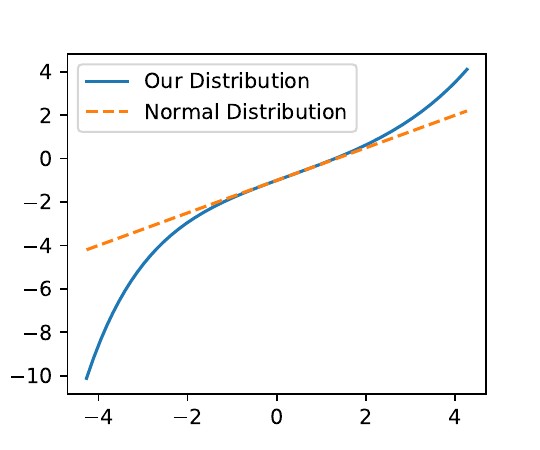}}
			\caption{{\footnotesize Q-Q plots against standard Gaussian}}
		\end{subfigure}\hspace{1em}
		\begin{subfigure}[t]{.25\linewidth}
			\centerline{\includegraphics[width=\linewidth]{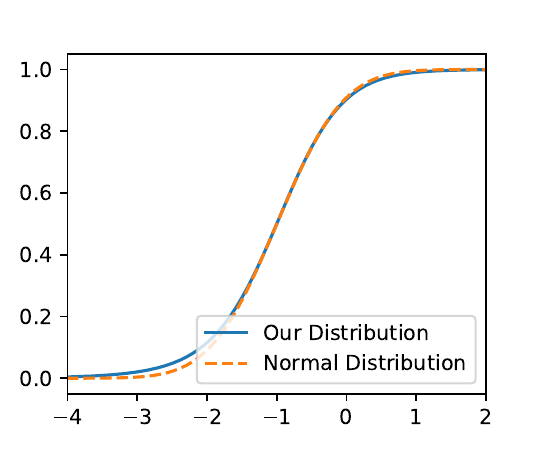}}
			\caption{{\footnotesize Cumulative functions}}
		\end{subfigure}\\
		\begin{subfigure}[t]{.25\linewidth}
			\centerline{\includegraphics[width=\linewidth]{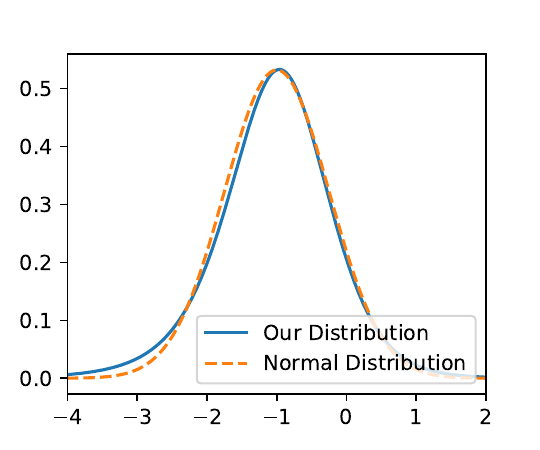}}
			\caption{{\footnotesize Density functions}}
		\end{subfigure}\hspace{1em}
		\begin{subfigure}[t]{.25\linewidth}
			\centerline{\includegraphics[width=\linewidth]{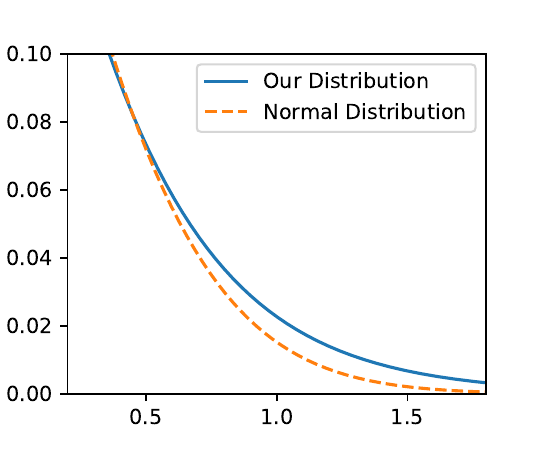}}
			\caption{{\footnotesize Right-tail densities}}
		\end{subfigure}\hspace{1em}
		\begin{subfigure}[t]{.25\linewidth}
			\centerline{\includegraphics[width=\linewidth]{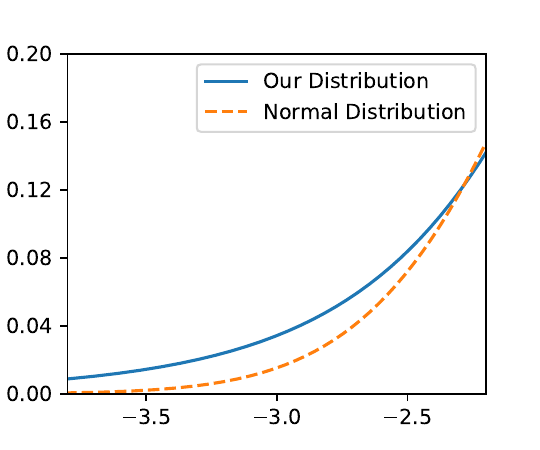}}
			\caption{{\footnotesize Left-tail densities}}
		\end{subfigure}
		\caption{The comparison between the generated distribution defined by the quantile function $f(\Phi^{-1}(\alpha))$ and the Gaussian $\mathcal{N}(\mu,\sigma^2(2/A+1)^2)$, where $f(x)=\mu+\sigma x(u^{x}/A+v^{-x}/A+1)$, with $g_1(x)=u^{x}/A$ and $g_2(x)=v^{-x}/A$. In this example, we set $A=4$, $\mu=-1$, $\sigma=0.5$, $u=1.5$, and $v=1.8$. The Gaussian distribution corresponds to the special case of $f$ where $u=v=1$.}
		\label{fig:exampleQQplot}
	\end{center}
\end{figure}
To gain a deeper understanding of the proposed transformation, we re-examine the distribution defined in Equation \eqref{htqf} (or equivalently, Equation \eqref{htqf.generative}). The corresponding transformation is given by
$f(x) = \mu + \sigma x \left(\frac{u^{x}}{A} + \frac{v^{-x}}{A} + 1\right)$, where $g_1(x) = \frac{u^{x}}{A}$ and $g_2(x) = \frac{v^{-x}}{A}$. We set $A = 4$, $\mu = -1$, $\sigma = 0.5$, $u = 1.5$, and $v = 1.8$. It can be easily verified that this function $f(x)$ satisfies the four conditions stated in Proposition \ref{prop:prop1}. The new distribution determined by $f(\Phi^{-1}(\alpha))$ yields a Q–Q plot against $\Phi^{-1}(\alpha)$ that is exactly the function graph of $f(x)$, as shown in Figure \ref{fig:exampleQQplot}. The corresponding cumulative distribution function, probability density function, and tail-side densities are also illustrated in Figure \ref{fig:exampleQQplot}. For comparison, we include a special case obtained by setting $u = v = 1$ in $f(x)$, which reduces to a Gaussian distribution $\mathcal{N}\left(\mu, \sigma^2(2/A + 1)^2\right)$. The results for this case are presented in the same figure. By comparing the tail-side densities of the two distributions, it is intuitively clear that the new distribution exhibits heavier tails than the Gaussian distribution on both sides.

\begin{figure}[t]
	\begin{center}
		\begin{subfigure}[t]{.25\linewidth}
			\centerline{\includegraphics[width=\linewidth]{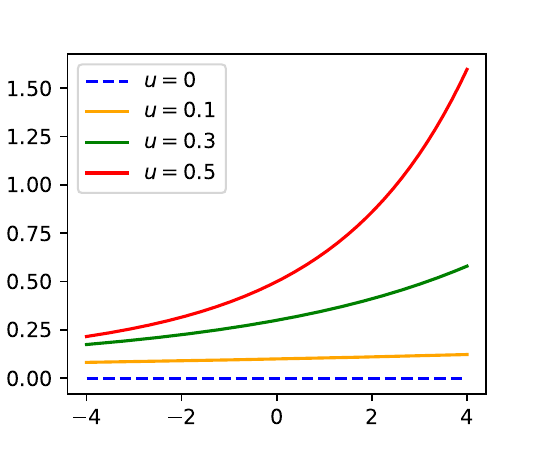}}
			\caption{{\footnotesize $g_1(x)$ under different values of $u$}}
		\end{subfigure}\hspace{1em}
		\begin{subfigure}[t]{.25\linewidth}
			\centerline{\includegraphics[width=\linewidth]{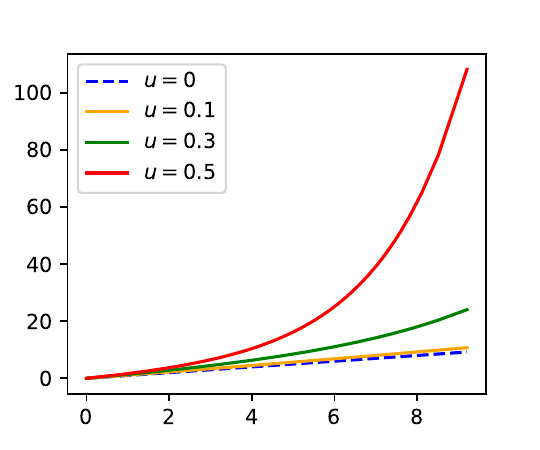}}
			\caption{{\footnotesize Q-Q plots of the generated distributions against standard exponential}}
		\end{subfigure}\hspace{1em}
		\begin{subfigure}[t]{.25\linewidth}
			\centerline{\includegraphics[width=\linewidth]{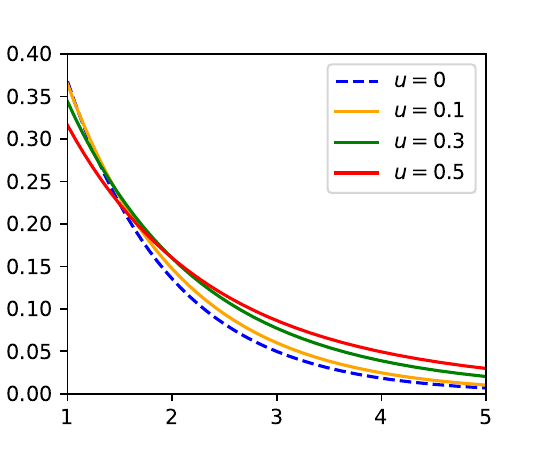}}
			\caption{{\footnotesize Right-tail densities of the generated distributions}}
		\end{subfigure}
		\caption{The comparison of the generated distributions defined by the quantile function $f(F^{-1}(\alpha))$, where $f(x) = x\left(\frac{e^{ux} - 1}{x} + 1\right)$, $g_1(x) = \frac{e^{ux} - 1}{x}$, and $g_2(x) = 0$. Here, $F$ denotes the cumulative distribution function of the standard exponential distribution, and $u$ takes values in ${0, 0.1, 0.3, 0.5}$. When $u = 0$, the transformation reduces to the standard exponential distribution itself.}
		\label{fig:exampleQQplot_exp}
	\end{center}
\end{figure}
When modeling the occurrence of an event, the exponential distribution is often the natural first choice. For this class of distributions, whose support covers only one half of $\mathbb{R}$, our framework can also be readily applied. To make the right tail heavier, we may simply set $g_2(x) \equiv 0$ and select an appropriate $g_1(x)$ in $f(x)$. In this example, we apply the transformation
\begin{equation}
	f(x) = x\left(\frac{e^{ux} - 1}{x} + 1\right),
\end{equation}
where $g_1(x) = (e^{ux} - 1)/x$, to the standard exponential distribution and examine how it modifies the right-tail behavior. We set $u = 0, 0.1, 0.3, 0.5$, and present the resulting distributions in Figure \ref{fig:exampleQQplot_exp}. As $g_1(x)$ goes to infinity more rapidly, the upper tail in the Q–Q plot against the standard exponential distribution becomes heavier, as does the right tail of the density. This provides empirical evidence that the parameter $u$ effectively governs the right-tail heaviness of the transformed distribution.

\begin{figure}[t]
	\begin{center}
		\begin{subfigure}[t]{.25\linewidth}
			\centerline{\includegraphics[width=\linewidth]{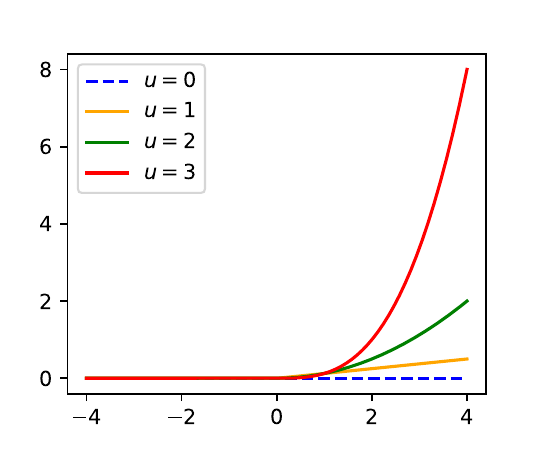}}
			\caption{{\footnotesize $g_1(x)$ under different values of $u$}}
		\end{subfigure}\hspace{1em}
		\begin{subfigure}[t]{.25\linewidth}
			\centerline{\includegraphics[width=\linewidth]{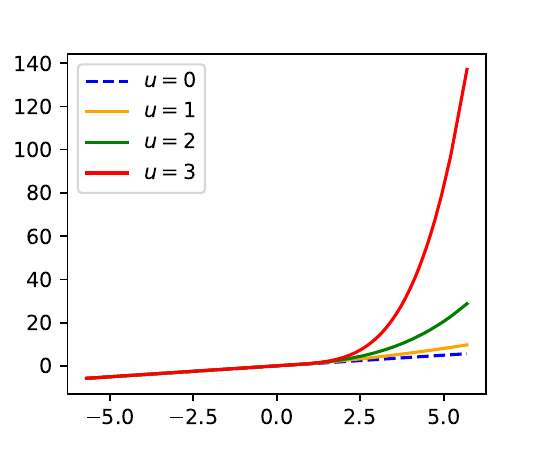}}
			\caption{{\footnotesize Q-Q plots of the generated distributions against the base distribution}}
		\end{subfigure}\hspace{1em}
		\begin{subfigure}[t]{.25\linewidth}
			\centerline{\includegraphics[width=\linewidth]{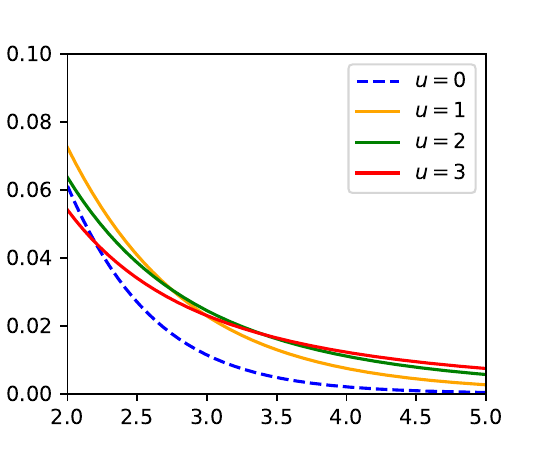}}
			\caption{{\footnotesize Right-tail densities of the generated distributions}}
		\end{subfigure}
		\caption{The comparison of the generated distributions defined by the quantile function $f(F^{-1}(\alpha))$, where $f(x) = x\left(g_1(x) + 1\right)$, $g_1(x) = \mathbb{I}_{\{x\ge 0\}}\cdot x^u/A$, and $g_2(x) = 0$. Here, $F$ denotes the cumulative distribution function of the base $t$-distribution with 10 degrees of freedom, $A$ is set to 8, and $u$ takes values in $\{1, 2, 3\}$. We denote the base distribution itself as $u=0$.}
		\label{fig:exampleQQplot_t}
	\end{center}
\end{figure}
The two previous examples both employed an exponential form for $g_1(x)$ or $g_2(x)$. Here, we present a third example using a power function form. Suppose a $t$-distribution with 10 degrees of freedom is chosen as the base distribution, with cumulative distribution function denoted by $F(x)$. For symmetry, we focus solely on the effect of $g_1(x)$, setting $g_2(x) \equiv 0$ and defining $g_1(x) = \mathbb{I}_{\{x\ge 0\}}\cdot x^u / A$. The transformation $f(x) = x(g_1(x) + 1)$ is applied to $F^{-1}(\alpha)$, with $u$ taking values from the set $\{1, 2, 3\}$. The resulting distributions are compared in Figure \ref{fig:exampleQQplot_t}, which also includes the base distribution itself as a reference, corresponding to the special case $g_1(x) \equiv 0$ for all $x \in \mathbb{R}$. From this comparison, it is evident that the transformation $f$ produces a new distribution with heavier tails, where the degree of heaviness is controlled by the parameter $u$. In addition to this empirical analysis, we provide rigorous theoretical results in the following.


\subsection{Tail Heaviness}

The tail heaviness of the constructed distribution depends on both the tail heaviness of the base distribution and the choice of functions $g_1(x)$ and $g_2(x)$. Owing to their symmetry, it suffices to consider the one-sided case as $x \to +\infty$ and to examine how the asymptotic behavior of $g_1(x)$ influences the right-tail heaviness of the new distribution (since $g_2(x) \to 0$). For convenience, we assume that the cumulative distribution function of the base distribution satisfies $F(x) < 1$ for any large $x$. Throughout, $f(x)$ is assumed to satisfy the four conditions stated in Proposition \ref{prop:prop1}. We begin with several specific cases, with all proofs provided in the appendix.

\begin{proposition} \label{prop:example1}
	Suppose the base distribution has a right-tail index $\nu$; equivalently, there exists a positive constant $\nu$ such that $\lim_{x \to +\infty} x^{\nu} \bar{F}(x) = c > 0$. If $\lim_{x \to +\infty} x^{-\nu'} g_1(x) = c' > 0$ for some positive $\nu'$, then the distribution determined by the quantile function $f(F^{-1}(\alpha))$ has a right-tail index $\nu / (1 + \nu')$. Alternatively, if $\lim_{x \to +\infty} e^{-t x^{\nu'}} g_1(x) = c' > 0$ for some $t > 0$ and $\nu' > 0$, then the distribution determined by $f(F^{-1}(\alpha))$ will exhibit a right tail as heavy as $(\log x)^{-\nu / \nu'}$.
\end{proposition}

Many distribution families satisfy the above characterization of $F(x)$, such as the Student's $t$ and Pareto distributions. Through the transformation $f$, a new distribution $F(f^{-1}(x))$ with more flexible tail behaviors can be constructed. However, other commonly used benchmark distributions, such as the normal and exponential, exhibit different asymptotic forms—specifically, the normal distribution satisfies $\lim_{x \to +\infty} x e^{x^2 / 2} \bar{F}(x) = c > 0$, while the exponential distribution satisfies $\lim_{x \to +\infty} e^{t x} \bar{F}(x) = c > 0$. For these cases, we present the following results.
\begin{proposition}
	\label{prop:example2}
(i) Suppose the distribution function $F$ exhibits a Gaussian tail, i.e., $\bar{F}(x) \sim x^{-1} e^{-x^2 / 2}$ as $x \to +\infty$, and we choose $g_1(x)$ such that $g_1(x) \sim x^{\nu - 1} e^{\nu x^2 / 2}$ for some $\nu > 0$ as $x \to +\infty$. Then, the distribution determined by $f(F^{-1}(\alpha))$ has a tail index of $1 / \nu$.\\
(ii) Suppose $F$ exhibits an exponential tail, i.e., there exists $t > 0$ such that $\bar{F}(x) \sim e^{-t x}$ as $x \to +\infty$, and we choose $g_1(x)$ such that $g_1(x) \sim x^{-1} e^{\nu x}$ for some $\nu > 0$ as $x \to +\infty$. Then, the distribution determined by $f(F^{-1}(\alpha))$ has a tail index of $t / \nu$.
\end{proposition}

	

The above examples illustrate specific cases of selecting the base distribution $F(x)$ and the transformation $f(x)$ (particularly the function $g_1(x)$). Next, we present two general results regarding the relationship between the choices of $F(x)$ and $f(x)$, and the resulting tail heaviness of the transformed distribution.
\begin{proposition}
	\label{prop:example4}
	Suppose $F$ is the base distribution function, and in the transformation $f(x)$, we have $\lim_{x \to +\infty} g_1(x) = +\infty$. If $F$ satisfies either of the following two conditions:\\
		\indent a) There exists $\rho < 0$ such that, for all $t > 0$, $\lim_{x \to +\infty} \bar{F}(t x) / \bar{F}(x) = t^{\rho}$; equivalently, $\bar{F}$ is regularly varying but not slowly varying;\\
		\indent b) There exists $t > 1$ such that $\lim_{x \to +\infty} \bar{F}(t x) / \bar{F}(x) = 0$;\\
	then the distribution determined by the quantile function $f(F^{-1}(\alpha))$ possesses a heavier right tail than the base distribution $F$.
\end{proposition} 
\begin{remark}
	The distribution derived in Equation \eqref{htqf} or \eqref{htqf.generative} exhibits a heavier right tail than the Gaussian distribution, since the Gaussian cumulative function satisfies Condition b) and $u^x / A \to +\infty$. A similar conclusion holds for the left tail. On the other hand, for any $n \ge 1$, the $n$-th moment of this distribution exists, as it is straightforward to verify the finiteness of $\mathbb{E}[Z^n (u^Z/A + v^{-Z}/A + 1)^n]$ when $Z$ follows the standard Gaussian distribution. Therefore, while the distribution's tails are heavier than Gaussian, they are not excessively heavy.
\end{remark}

Notice that a wide range of distributions satisfy either of the two conditions in Proposition \ref{prop:example4}. Consequently, this proposition implies that for many choices of $F$, any transformation $f(x)$ satisfying $g_1(x)\to +\infty$ will produce a new distribution with a heavier tail. Moreover, as further illustrated by the following proposition, by appropriately specifying $g_1(x)$ and $g_2(x)$, our proposed transformation can generate a new distribution $F_2$ whose tail can be made arbitrarily heavier than that of the base distribution $F_1$.
\begin{proposition} \label{theorem:rev_theorem1}
	Let $F_1$ and $F_2$ be two distribution functions, where $F_2$ has a heavier right tail than $F_1$, that is,
	$\lim_{x\to +\infty}\bar{F}_2(x)/\bar{F}_1((x-\mu)/\sigma)=+\infty, \forall \mu\in\mathbb{R}, \sigma>0$.
	Suppose further that $F_2$ admits a decreasing density function $p_2$ for sufficiently large $x$, and that $F_2^{-1}(\alpha)/F_1^{-1}(\alpha)$ is non-decreasing as $\alpha\to 1^{-}$. Then there exist functions $g_1(x)$ and $g_2(x)$ such that the transformation $f(x)$ defined in Equation~(\ref{eqn:eqn_f}) yields a distribution determined by the quantile function $f(F_1^{-1}(\alpha))$ whose right-tail heaviness is identical to that of $F_2$.
\end{proposition}
Detailed proofs of these propositions are provided in the appendix.

\section{Parameter Estimation}
\label{sec:parameter}

Apart from its simplicity in sampling, another major advantage of our method over traditional density-based distribution modeling approaches lies in its computational convenience for optimization. The mathematical formulation of our method is substantially more concise than that of density-based approaches. In particular, by adopting a simple parametric form for $g_1(x)$ (or $g_2(x)$), denoted as $g_1(x;\theta_1)$ (or $g_2(x;\theta_2)$), the computation of the derivative $\partial f / \partial \theta_1$ becomes significantly easier than in density-based approaches, which typically require evaluating derivatives of complex special functions such as the beta or gamma function. In fact, all the examples presented in Section \ref{sec:examples} employ simple parametric forms for $g_1(x)$ or $g_2(x)$.

A direct implication of this advantage is the simplicity of parameter estimation. Suppose $F$ denotes the base distribution, and $f_\theta(F^{-1}(\alpha))$ represents the quantile function of the constructed distribution, where $f_\theta$ is a properly designed function parameterized by $\theta = \{\mu, \sigma, \theta_1, \theta_2\}$. Since the quantiles of the target random variable can be readily computed, parameter estimation can naturally be carried out via quantile regression \citep{koenker2001quantile,koenker2005quantile}, given observations $\{y_i\}_{i=1}^N$:
\begin{equation}
	\min_{\theta} \sum_{\alpha\in A}\frac{1}{N}\sum_{i=1}^N L_\alpha(y_i, f_\theta(F^{-1}(\alpha))),\quad L_\alpha(y,q) = (\alpha-\mathbb{I}_{\{y<q\}})(y-q),
\end{equation}
where $A$ denotes a set of probability levels, such as $A = \{0.01, 0.02, \dots, 0.98, 0.99\}$ with $|A| = 99$.

The quantile function $f_\theta(F^{-1}(\alpha))$ can also represent a conditional distribution, given certain predictors. Conditional distribution modeling is a central task in statistics, machine learning, and related fields. More formally, let $f_{\theta}(F^{-1}(\alpha))$ denote the conditional distribution $\mathbb{P}(Y \mid X = x)$, where $\theta = s(x)$ characterizes how the distribution depends on the predictors $x$. Under this conditional setting, parameter estimation—or equivalently, model training—can again be naturally conducted via quantile regression, since the conditional quantiles remain analytically tractable. Given observations $\{(x_i, y_i)\}_{i=1}^N$, the function $s$ can be learned by solving
\begin{equation}
	\min_{s\in\mathcal{H}} \sum_{\alpha\in A}\frac{1}{N}\sum_{i=1}^N L_\alpha(y_i, f_{s(x_i)}(F^{-1}(\alpha))),
\end{equation}
where $\mathcal{H}$ denotes the hypothesis space of candidate functions.
Traditional density-based approaches encounter significant challenges when employed in machine learning, particularly in deep learning, which typically relies on large-scale gradient-based optimization \citep{kingma2014adam}.

\section{Numerical Results}
\label{sec:applications}

\begin{table}[t]
	\centering
	\caption{The percentage of stocks for which the null hypothesis of chi-square test is rejected at the 5\% significance level, as obtained from each model across daily, weekly, and monthly log-returns (lower values indicate a better fit).}
	\label{tab:chi_square}
	\begin{tabular}{p{4cm} p{2cm} p{2cm} p{2cm}}
		\hline 
		Distributions & Daily & Weekly & Monthly \\
		\hline 
		\textit{Five-parameter} & & & \\
		SGT & 9.9 & 7.4 & 14.8 \\
		\hline 
		\textit{Four-parameter} & & & \\
		GB2 & 8.5 & 9.9 & 14.1 \\
		ST & 11.6 & 7.0 & 11.6 \\
		PGML & 10.2 & 6.7 & 11.6 \\
		\hline 
		\textit{Three-parameter} & & & \\
		GG & 52.5 & 98.6 & 29.6 \\
		T & 8.1 & 9.9 & 14.4 \\
		\hline 
		\textit{Two-parameter} & & & \\
		N & 44.7 & 98.2 & 23.2 \\
		L & 19.0 & 68.3 & 44.4 \\
		\hline 
	\end{tabular}
\end{table}

\begin{table}[t]
	\centering
	\caption{Mean and standard error of the Kolmogorov–Smirnov measure across all stocks. Smaller values indicate a better goodness of fit.}
	\label{tab:KS-measures}
	\begin{tabular}{p{4cm} p{2.5cm} p{2.5cm} p{2.5cm}}
		\hline 
		Distributions & Daily & Weekly & Monthly \\
		\hline 
		\textit{Five-parameter} & & & \\
		SGT & 0.031 (0.007) & 0.017 (0.004) & 0.038 (0.011) \\
		\hline 
		\textit{Four-parameter} & & & \\
		GB2 & 0.032 (0.007) & 0.019 (0.005) & 0.038 (0.008) \\
		ST & 0.034 (0.008) & 0.019 (0.004) & 0.039 (0.008) \\
		PGML & 0.033 (0.007) & 0.018 (0.004) & 0.036 (0.007) \\
		\hline 
		\textit{Three-parameter} & & & \\
		GG & 0.071 (0.022) & 0.075 (0.019) & 0.072 (0.027) \\
		T & 0.036 (0.008) & 0.021 (0.004) & 0.044 (0.010) \\
		\hline 
		\textit{Two-parameter} & & & \\
		N & 0.065 (0.018) & 0.066 (0.016) & 0.061 (0.019) \\
		L & 0.049 (0.013) & 0.036 (0.008) & 0.066 (0.015) \\
		\hline 
	\end{tabular}
\end{table}

\begin{table}[t]
	\centering
	\caption{Mean and standard error of the Kuiper measure across all stocks. Smaller values indicate a better goodness of fit.}
	\label{tab:kuiper-measures}
	\begin{tabular}{p{4cm} p{2.5cm} p{2.5cm} p{2.5cm}}
		\hline 
		Distributions & Daily & Weekly & Monthly \\
		\hline 
		\textit{Five-parameter} & & & \\
		SGT & 0.055 (0.010) & 0.031 (0.006) & 0.066 (0.013) \\
		\hline 
		\textit{Four-parameter} & & & \\
		GB2 & 0.058 (0.011) & 0.035 (0.007) & 0.070 (0.013) \\
		ST & 0.063 (0.013) & 0.034 (0.006) & 0.070 (0.013) \\
		PGML & 0.060 (0.012) & 0.033 (0.006) & 0.067 (0.012) \\
		\hline 
		\textit{Three-parameter} & & & \\
		GG & 0.117 (0.031) & 0.122 (0.031) & 0.107 (0.033) \\
		T & 0.065 (0.013) & 0.037 (0.007) & 0.078 (0.015) \\
		\hline 
		\textit{Two-parameter} & & & \\
		N & 0.116 (0.032) & 0.122 (0.031) & 0.106 (0.032) \\
		L & 0.079 (0.018) & 0.059 (0.012) & 0.105 (0.020) \\
		\hline 
	\end{tabular}
\end{table}

\begin{table}[t]
	\centering
	\caption{Mean and standard deviation of negative log-likelihood values across all stocks. Lower values indicating a better fit.}
	\label{tab:NLL}
	\resizebox{\textwidth}{!}{
		\begin{tabular}{p{3.5cm} p{3.5cm} p{3.5cm} p{3.5cm}}
			\hline 
			Distributions & Daily & Weekly & Monthly \\
			\hline 
			\textit{Five-parameter} & & & \\
			SGT & -689.286 (77.756) & -1531.435 (176.476) & -224.823 (44.173) \\
			\hline 
			\textit{Four-parameter} & & & \\
			GB2 & -687.456 (78.321) & -1523.719 (177.620) & -224.061 (44.696) \\
			ST & -688.350 (77.714) & -1530.740 (176.440) & -224.419 (44.211) \\
			PGML & -685.520 (75.404) & -1529.090 (176.010) & -223.769 (44.318) \\
			\hline 
			\textit{Three-parameter} & & & \\
			GG & -678.598 (77.908) & -1470.544 (179.337) & -219.980 (46.418) \\
			T & -687.982 (77.737) & -1529.552 (176.076) & -223.481 (44.003) \\
			\hline 
			\textit{Two-parameter} & & & \\
			N & -678.114 (77.801) & -1466.549 (179.226) & -218.799 (46.770) \\
			L & -685.847 (78.038) & -1519.736 (175.062) & -219.044 (43.642) \\
			\hline 
	\end{tabular}}
\end{table}

\begin{table}[t]
	\centering
	\caption{Frequency with which each distribution ranks among the top two based on NLL.}
	\label{tab:NLL-rank}
	\begin{tabular}{p{4cm} p{2cm} p{2cm} p{2cm}}
		\hline 
		Distributions & Daily & Weekly & Monthly \\
		\hline 
		\textit{Five-parameter} & & & \\
		SGT & 280 & 278 & 253 \\
		\hline
		\textit{Four-parameter} & & & \\
		GB2 & 166 & 64 & 116 \\
		ST & 92 & 187 & 137 \\
		PGML & 3 & 37 & 47 \\
		\hline
		\textit{Three-parameter} & & & \\
		GG & 0 & 0 & 6 \\
		T & 0 & 1 & 1 \\
		\hline
		\textit{Two-parameter} & & & \\
		N & 1 & 0 & 7 \\
		L & 26 & 1 & 1 \\
		\hline 
	\end{tabular}
\end{table}

%

In this section, we evaluate our proposed method on stock return modeling and compare its performance with several widely used distributions. Our framework allows for numerous choices of $f(x)$, providing great flexibility in the resulting distributions. To maintain clarity and conciseness in the numerical analysis, we focus on the specific formulation given in Equation \eqref{htqf} (or equivalently, Equation \eqref{htqf.generative}), which has been validated in previous studies. We refer to this distribution as PGML (Parsimonious GML).

Because a wide variety of distributions have been applied in finance, we evaluate PGML against seven widely used traditional distributions: the skewed generalized $t$ (SGT), generalized beta of the second kind (GB2), skewed $t$ (ST), generalized gamma (GG) \citep{stacy1962generalization}, Student's $t$ (T), normal (N), and Laplace (L) distributions. Among these, SGT is a five-parameter distribution; GB2 and ST are four-parameter distributions; GG and T are three-parameter distributions; and N and L are two-parameter distributions. The PGML model in this paper is a four-parameter distribution.

All models are applied to the daily, weekly, and monthly log-returns of 284 S\&P 500 stocks over the period 2010–2024. Weekly and monthly log-returns are computed using the entire fifteen-year sample, while daily log-returns are based on data from January 4 to December 31, 2010. Since log-returns can take both positive and negative values, if a distribution's support is $\mathbb{R}^+$, we apply a logarithmic transformation to the corresponding random variable to obtain a new distribution whose support is $\mathbb{R}$.


For the seven traditional distributions, parameters are estimated via maximum likelihood. Goodness-of-fit is evaluated using the chi-square test, the Kolmogorov–Smirnov and Kuiper statistics (both based on the maximum distance between the empirical and fitted cumulative distribution functions), as well as the negative log-likelihood (NLL).

\subsection{Chi‐Square Test}

The chi-square goodness-of-fit test begins by trimming the most extreme $\alpha$ percent of observations from both tails, and then partitioning the remaining range into $b$ equally spaced bins (with $\alpha = 0.05$ and $b = 10$ in our setting). For each bin, the observed frequency $E_i$ is compared with the expected frequency $T_i$ under the fitted distribution, and the test statistic is computed as
\begin{equation*}
	\chi^2 = \sum_{i=1}^{b+2} \frac{(E_i - T_i)^2}{T_i},
\end{equation*}
which, under the null hypothesis, asymptotically follows a chi-square distribution with $b - p + 1$ degrees of freedom, where $p$ denotes the number of estimated parameters.

Table \ref{tab:chi_square} reports the percentage of stocks for which the null hypothesis is rejected at the 5\% significance level, as obtained from each model across daily, weekly, and monthly log-returns (lower values indicate a better fit). The PGML model demonstrates competitive performance for daily returns (10.2\%), closely comparable to other flexible four- and five-parameter distributions such as GB2 (8.5\%) and SGT (9.9\%). More importantly, PGML achieves the best fit for weekly (6.7\%) and monthly (11.6\%) returns, outperforming all other candidate distributions, including the five-parameter SGT. In contrast, simpler distributions such as GG, N, and L exhibit very high rejection rates—exceeding 60\% for weekly data—indicating their poor fit. Although the T distribution performs moderately well, it is still outperformed by PGML for both weekly and monthly data.

\subsection{Kolmogorov–Smirnov and Kuiper Tests}

Compared to the chi-square test's reliance on arbitrary binning and its bias in sparse tail regions, the Kolmogorov–Smirnov test directly measures the maximum deviation between the empirical distribution function and the theoretical cumulative distribution function, thus avoiding subjective bin selection and providing a more robust goodness-of-fit assessment. However, it is less sensitive to differences in the distribution tails, and its critical values typically assume known parameters, which may reduce accuracy when parameters are estimated from the sample. To overcome this limitation, the Kuiper test introduces a symmetric statistic, offering balanced sensitivity across the distribution support. The Kolmogorov–Smirnov measure ($m_{\text{KS}}$)
and Kuiper measure ($m_{\text{K}}$) are defined as follows,
\begin{align*}
	& D_+ = \max_{1 \leq i \leq n}\left( \frac{i}{n} - F(y_i) \right),\quad \ D_- = \max_{1 \leq i \leq n}\left( F(y_i) - \frac{i-1}{n} \right), \\ 
	& m_{\text{KS}} = \max\left( D_+, D_- \right),\\
	& m_{\text{K}} = D_+ + D_-,
\end{align*}
where the data $\{y_i\}_{i=1}^n$ are ordered from 1 to $n$, and $F(y)$ denotes the theoretical cumulative distribution function. In both tests, smaller statistic values indicate a better goodness of fit.


Tables~\ref{tab:KS-measures} and~\ref{tab:kuiper-measures} report the mean and standard error of $m_{\text{KS}}$ and $m_{\text{K}}$ across all stocks, respectively. As expected, the more flexible four- and five-parameter distributions (SGT, GB2, ST, and PGML) outperform the simpler models (GG, T, N, and L), which exhibit substantially larger values. Within this context, the PGML model demonstrates highly competitive performance. For daily returns, its $m_{\text{KS}}$ (0.033) and $m_{\text{K}}$ (0.060) closely match those of SGT and GB2. More importantly, for weekly and monthly returns, PGML ranks either first or a close second across both metrics, achieving $m_{\text{KS}}$ values of 0.018 and 0.036, and $m_{\text{K}}$ values of 0.033 and 0.067, respectively—underscoring its strong tail-fitting capability. Overall, PGML consistently matches or surpasses the benchmark models across all time horizons.

\subsection{Negative Log-Likelihood}


The negative log-likelihood (NLL) assesses the goodness of fit by computing the negative average log-probability of the data under the estimated distribution, with lower values indicating a better fit. Table~\ref{tab:NLL} reports the mean and standard deviation of NLL values across all stocks. PGML performs slightly worse than GB2 and ST, though the differences are relatively small, particularly for monthly returns.
Table~\ref{tab:NLL-rank} further summarizes how frequently each distribution ranks among the top two based on NLL over all stocks. PGML attains top-two rankings 3, 37, and 47 times for daily, weekly, and monthly data, respectively—less frequently than competitors such as SGT and GB2. Overall, simpler models perform considerably worse than the more flexible four- and five-parameter distributions.


\section{Conclusion}
\label{sec:conclusion}

This paper introduces a novel generative machine learning approach for modeling the prevalent non-Gaussian heavy tails observed in financial data. Departing significantly from traditional methods that rely on complex density functions like the four-parameter GB2 and the five-parameter SGT, the proposed method constructs new distributions by applying a parsimonious nonlinear transformation to a simple base random variable (e.g., Gaussian). This approach yields an explicit quantile function, combining high flexibility with mathematical elegance and computational convenience.

We propose a generalized transformation form, meticulously designed to flexibly control tail heaviness and asymmetry. A key theoretical contribution is the rigorous derivation of the tail properties of the resulting distribution relative to the base distribution. The analysis proves that by appropriately designing the transformation, one can generate distributions with arbitrarily heavy tails, with precise characterizations of tail heaviness.
For parameter estimation, the explicit quantile function naturally lends itself to quantile regression. This estimation strategy avoids the gradient computation difficulties often encountered in traditional maximum likelihood estimation of complex densities. Furthermore, it facilitates seamless extension to conditional distribution modeling, making the method particularly suitable for machine learning tasks that rely on gradient-based optimization.

In numerical experiment, we compare the specific PGML model against a range of traditional distributions, including SGT, GB2, GG, and others. The models are assessed on daily, weekly, and monthly stock log-returns using multiple metrics: the chi-square test, Kolmogorov-Smirnov statistic, Kuiper statistic, and negative log-likelihood. The results demonstrate the robust and competitive performance of the PGML model. It performs well on daily returns and achieves top-tier performance, often ranking first or second, on weekly and monthly data, underscoring its effectiveness in capturing tail risk across different time horizons.

In conclusion, the framework developed in this paper offers a fundamentally new and powerful paradigm for heavy-tail modeling. Its core strengths are multifaceted: the model structure is parsimonious, facilitating easy sampling and computation; the tail properties are explicitly derived and controllable; and parameter estimation is robust and highly compatible with modern machine learning workflows. Consequently, this method holds significant promise not only for direct applications in finance but also as a versatile tool for conditional distribution prediction in broader machine learning contexts.


\vspace{5em}

\bibliographystyle{elsarticle-harv} 
\bibliography{Bibliography-MM-MC}       

@incollection{mcdonald2008some,
  title={Some generalized functions for the size distribution of income},
  author={McDonald, James B},
  booktitle={Modeling Income Distributions and Lorenz Curves},
  pages={37--55},
  year={2008},
  publisher={Springer}
}

@article{kingma2019introduction,
  title={An introduction to variational autoencoders},
  author={Kingma, Diederik P and Welling, Max},
  journal={Foundations and Trends{\textregistered} in Machine Learning},
  volume={12},
  number={4},
  pages={307--392},
  year={2019},
  publisher={Now Publishers, Inc.}
}

@inproceedings{arjovsky2017wasserstein,
  title={Wasserstein generative adversarial networks},
  author={Arjovsky, Martin and Chintala, Soumith and Bottou, L{\'e}on},
  booktitle={International Conference on Machine Learning},
  pages={214--223},
  year={2017},
  organization={PMLR}
}

@article{kobyzev2020normalizing,
  title={Normalizing flows: An introduction and review of current methods},
  author={Kobyzev, Ivan and Prince, Simon JD and Brubaker, Marcus A},
  journal={IEEE Transactions on Pattern Analysis and Machine Intelligence},
  volume={43},
  number={11},
  pages={3964--3979},
  year={2020},
  publisher={IEEE}
}

@inproceedings{rombach2022high,
  title={High-resolution image synthesis with latent diffusion models},
  author={Rombach, Robin and Blattmann, Andreas and Lorenz, Dominik and Esser, Patrick and Ommer, Bj{\"o}rn},
  booktitle={Proceedings of the IEEE/CVF Conference on Computer Vision and Pattern Recognition},
  pages={10684--10695},
  year={2022}
}

@article{yang2023diffusion,
  title={Diffusion models: A comprehensive survey of methods and applications},
  author={Yang, Ling and Zhang, Zhilong and Song, Yang and Hong, Shenda and Xu, Runsheng and Zhao, Yue and Zhang, Wentao and Cui, Bin and Yang, Ming-Hsuan},
  journal={ACM Computing Surveys},
  volume={56},
  number={4},
  pages={1--39},
  year={2023},
  publisher={ACM New York, NY, USA}
}

@article{croitoru2023diffusion,
  title={Diffusion models in vision: A survey},
  author={Croitoru, Florinel-Alin and Hondru, Vlad and Ionescu, Radu Tudor and Shah, Mubarak},
  journal={IEEE Transactions on Pattern Analysis and Machine Intelligence},
  volume={45},
  number={9},
  pages={10850--10869},
  year={2023},
  publisher={Ieee}
}

@article{koenker2001quantile,
  title={Quantile regression},
  author={Koenker, Roger and Hallock, Kevin F},
  journal={Journal of Economic Perspectives},
  volume={15},
  number={4},
  pages={143--156},
  year={2001},
  publisher={American Economic Association}
}

@book{koenker2005quantile,
  title={Quantile regression},
  author={Koenker, Roger},
  volume={38},
  year={2005},
  publisher={Cambridge university press}
}

@article{lecun2015deep,
  title={Deep learning},
  author={LeCun, Yann and Bengio, Yoshua and Hinton, Geoffrey},
  journal={Nature},
  volume={521},
  number={7553},
  pages={436--444},
  year={2015},
  publisher={Nature Publishing Group UK London}
}

@article{stacy1962generalization,
  title={A generalization of the gamma distribution},
  author={Stacy, Edney W},
  journal={The Annals of Mathematical Statistics},
  pages={1187--1192},
  year={1962},
  publisher={JSTOR}
}

@article{theodossiou1998financial,
  title={Financial data and the skewed generalized t distribution},
  author={Theodossiou, Panayiotis},
  journal={Management Science},
  volume={44},
  number={12-part-1},
  pages={1650--1661},
  year={1998},
  publisher={INFORMS}
}

@article{cont2001empirical,
  title={Empirical properties of asset returns: stylized facts and statistical issues},
  author={Cont, Rama},
  journal={Quantitative Finance},
  volume={1},
  number={2},
  pages={223},
  year={2001},
  publisher={IOP Publishing}
}

@book{mcneil2015quantitative,
  title={Quantitative risk management: concepts, techniques and tools-revised edition},
  author={McNeil, Alexander J and Frey, R{\"u}diger and Embrechts, Paul},
  year={2015},
  publisher={Princeton university press}
}

@book{rachev2003handbook,
  title={Handbook of heavy tailed distributions in finance: Handbooks in finance, Book 1},
  author={Rachev, Svetlozar Todorov},
  year={2003},
  publisher={Elsevier}
}

@article{yan2019cross,
  title={Cross-sectional learning of extremal dependence among financial assets},
  author={Yan, Xing and Wu, Qi and Zhang, Wen},
  journal={Advances in Neural Information Processing Systems},
  volume={32},
  year={2019}
}

@book{jondeau2007financial,
  title={Financial modeling under non-Gaussian distributions},
  author={Jondeau, Eric and Poon, Ser-Huang and Rockinger, Michael},
  year={2007},
  publisher={Springer Science \& Business Media}
}

@article{yan2018parsimonious,
  title={Capturing deep tail risk via sequential learning of quantile dynamics},
  author={Wu, Qi and Yan, Xing},
  journal={Journal of Economic Dynamics and Control},
  volume={109},
  pages={103771},
  year={2019},
  publisher={Elsevier}
}

@inproceedings{kingma2014adam,
  title={Adam: a method for stochastic optimization},
  author={Kingma, Diederik P and Ba, Jimmy},
  booktitle={International Conference on Learning Representations},
  year={2014}
}

@inproceedings{kingma2013auto,
  title={Auto-encoding variational bayes},
  author={Kingma, Diederik P and Welling, Max},
  booktitle={International Conference on Learning Representations},
  year={2014}
}

@article{GAN,
  title={Generative adversarial nets},
  author={Goodfellow, Ian and Pouget-Abadie, Jean and Mirza, Mehdi and Xu, Bing and Warde-Farley, David and Ozair, Sherjil and Courville, Aaron and Bengio, Yoshua},
  journal={Advances in Neural Information Processing Systems},
  volume={27},
  year={2014}
}

@article{NormalizingFlow,
  title={Normalizing flows for probabilistic modeling and inference},
  author={Papamakarios, George and Nalisnick, Eric and Rezende, Danilo Jimenez and Mohamed, Shakir and Lakshminarayanan, Balaji},
  journal={Journal of Machine Learning Research},
  volume={22},
  number={1},
  pages={2617--2680},
  year={2021},
  publisher={JMLRORG}
}



\vspace{5em}

\appendix
\section{Some Proofs}
\label{AppendixA}
\addcontentsline{toc}{section}{Appendices}
\renewcommand{\thesubsection}{\Alph{subsection}}

In this appendix, we present the proofs of some propositions introduced in this paper.

\textbf{For Proposition \ref{prop:example1}:}
\begin{proof}
	For the distribution determined by $f(F^{-1}(\alpha))$, the cumulative distribution function is $F(f^{-1}(x))$. Then, we have
	\begin{align*}
		\lim_{x\to+\infty}x^{\frac{\nu}{1+\nu'}}\left(1-F(f^{-1}(x))\right) = &
		\lim_{x\to+\infty}f(x)^{\frac{\nu}{1+\nu'}}\bar{F}(x) \\
		= & \lim_{x\to+\infty}c\left(\frac{f(x)}{x}\right)^{\frac{\nu}{1+\nu'}}x^{\frac{-\nu\nu'}{1+\nu'}} \\
		= &
		\lim_{x\to+\infty}c\left(\frac{\frac{\mu}{x}+\sigma(g_1(x)+g_2(x)+1)}{x^{\nu'}}\right)^{\frac{\nu}{1+\nu'}} \\
		= & ~ c(\sigma c')^{\frac{\nu}{1+\nu'}}.
	\end{align*}
	Alternatively,
	\begin{align*}
		\lim_{x\to+\infty}(\log x)^{\frac{\nu}{\nu'}}\left(1-F(f^{-1}(x))\right) = &
		\lim_{x\to+\infty}(\log f(x))^{\frac{\nu}{\nu'}}\bar{F}(x) \\
		= & \lim_{x\to+\infty}c\left(\log \frac{f(x)}{x}+\log x\right)^{\frac{\nu}{\nu'}}x^{-\nu}  \\
		= & \lim_{x\to+\infty}c\left(\frac{\log(\sigma c')+t x^{\nu'}+\log x}{x^{\nu'}}\right)^{\frac{\nu}{\nu'}} \\
		= & ~ c t^{\frac{\nu}{\nu'}}.
	\end{align*}
\end{proof}

\textbf{For Proposition \ref{prop:example2}:}
\begin{proof}
	For (i),
	\begin{align*}
		\lim_{x\to+\infty}x^{\frac{1}{\nu}}\left(1-F(f^{-1}(x))\right) = &
		\lim_{x\to+\infty}f(x)^{\frac{1}{\nu}}\bar{F}(x) \\
		= & \lim_{x\to+\infty}c\left(\frac{f(x)}{\sigma xg_1(x)}\right)^{\frac{1}{\nu}}\frac{\left(\sigma xg_1(x)\right)^{\frac{1}{\nu}}}{xe^{x^2/2}} \\
		= & \lim_{x\to+\infty}c(\sigma c')^{\frac{1}{\nu}} \frac{\left(xx^{\nu-1}e^{\nu x^2/2}\right)^{\frac{1}{\nu}}}{xe^{x^2/2}} \\
		= & ~ c(\sigma c')^{\frac{1}{\nu}},
	\end{align*}
	where $c$ and $c'$ are the corresponding constants. For (ii),
	\begin{align*}
		\lim_{x\to+\infty}x^{\frac{t}{\nu}}\left(1-F(f^{-1}(x))\right) = &
		\lim_{x\to+\infty}f(x)^{\frac{t}{\nu}}\bar{F}(x) \\
		= & \lim_{x\to+\infty}c \left(\frac{f(x)}{\sigma xg_1(x)}\right)^{\frac{t}{\nu}}\frac{(\sigma xg_1(x))^{\frac{t}{\nu}}}{e^{tx}} \\
		= & \lim_{x\to+\infty} c(\sigma c')^{\frac{t}{\nu}} \frac{(xx^{-1}e^{\nu x})^{\frac{t}{\nu}}}{e^{tx}} \\
		= & ~ c(\sigma c')^{\frac{t}{\nu}}.
	\end{align*}
\end{proof}


\textbf{For Proposition \ref{prop:example4}:}
\begin{proof}
	We will show that the following limit equals $+\infty$ for any $\mu \in \mathbb{R}$ and $\sigma > 0$:
	\begin{equation}
		\lim_{x\to +\infty}\frac{1-F(f^{-1}(x))}{1-F(\frac{x-\mu}{\sigma})}=\lim_{x\to +\infty}\frac{\bar{F}(x)}{\bar{F}(\frac{f(x)-\mu}{\sigma})}.
	\end{equation}
	Since arbitrary $\mu$ and $\sigma$ also appear in the formulation of $f(x)$, they can be absorbed, and the above limit simplifies to $\lim_{x\to +\infty}\bar{F}(x)/\bar{F}(f(x))$. Because $\lim_{x\to +\infty}f(x)/x = \lim_{x\to +\infty}\sigma (g_1(x) + 1) = +\infty$, for any large $t$, we have $f(x) > tx$ when $x$ is sufficiently large. When Condition a) holds,
	\begin{equation}
		\frac{\bar{F}(x)}{\bar{F}(f(x))}\ge \frac{\bar{F}(x)}{\bar{F}(tx)} \to t^{-\rho},\quad x\to +\infty.
	\end{equation}
	Thus, for sufficiently large $x$, we have $\bar{F}(x)/\bar{F}(f(x))\ge \bar{F}(x)/\bar{F}(tx) \ge t^{-\rho}-1$. Due to the arbitrariness of $t$, it follows that $\lim_{x\to +\infty}\bar{F}(x)/\bar{F}(f(x))=+\infty$.
	When Condition (b) holds, for the fixed $t>1$ and any large $x$,
	\begin{equation}
		\frac{\bar{F}(x)}{\bar{F}(f(x))}\ge \frac{\bar{F}(x)}{\bar{F}(tx)} \to +\infty,\quad x\to +\infty.
	\end{equation}
	Hence, the proof is complete.
\end{proof}

\textbf{For Proposition \ref{theorem:rev_theorem1}:} 
\begin{proof}
	We first prove that as $x\to +\infty$, $F_2^{-1}(F_1(x))/x \to +\infty$, which is equivalent to $\lim_{\alpha\to 1^{-}}F_2^{-1}(\alpha)/F_1^{-1}(\alpha)=+\infty$. Suppose this does not hold. Then there must exist some $M>0$ and a sequence $x_i\to +\infty$ such that $F_2^{-1}(F_1(x_i))\le Mx_i$. Consequently, $F_1(x_i)\le F_2(Mx_i)$, implying $\bar{F_1}(x_i)\ge \bar{F_2}(Mx_i)$, which contradicts the assumption of tail heaviness that $\bar{F_2}(Mx)/\bar{F_1}(x) \to +\infty$. Furthermore, $F_2^{-1}(F_1(x))/x$ is clearly non-decreasing for sufficiently large $x$.
	
	We define 
	\begin{equation}
	g_1(x)=\frac{-\mu+F_2^{-1}(F_1(x))}{\sigma x}-1,
	\end{equation}
	for sufficiently large $x$, where $\mu$ and $\sigma$ are arbitrary parameters in the formulation of $f(x)$. Clearly, $g_1(x)\to +\infty$ as $x\to +\infty$.
	Next, we set $g_2(x)$ to be a rapidly decaying function satisfying 
	\begin{equation}
	\lim_{x\to +\infty}xg_2(x)=0,\quad \text{and} \quad \lim_{x\to +\infty}xg_2(x)\cdot \frac{p_2(h(x))}{\bar{F_2}(h(x))}=0, 
	\end{equation}
	where $h(x)=F_2^{-1}(F_1(x))$ (for instance, one may simply take $g_2(x)\equiv 0$ for large $x$). It is straightforward to ensure that $g_1(x)$ and $g_2(x)$ satisfy all the conditions stated in Proposition \ref{prop:prop1}. Now,
	\begin{equation}
		\lim_{x\to+\infty}\frac{1-F_2(x)}{1-F_1(f^{-1}(x))} = \lim_{x\to+\infty}\frac{\bar{F_2}(f(x))}{\bar{F_1}(x)} = 
		\lim_{x\to+\infty}\frac{\bar{F_2}\left(\sigma xg_2(x)+h(x)\right)}{\bar{F_1}(x)}.
	\end{equation}
	Applying the mean value theorem and noting that $\bar{F_2}(h(x))=\bar{F_1}(x)$, we have
	\begin{equation}
	\begin{aligned}
		\frac{\bar{F_2}\left(\sigma xg_2(x)+h(x)\right)}{\bar{F_1}(x)}
		= & \frac{\bar{F_2}(h(x))-p_2(\xi_x)\sigma xg_2(x)}{\bar{F_2}(h(x))} \\
		= & 1-\sigma xg_2(x)\frac{p_2(\xi_x)}{\bar{F_2}(h(x))} \to 1,
	\end{aligned}
	\end{equation}
	because 
	\begin{equation}
		\sigma xg_2(x)\frac{p_2(\xi_x)}{\bar{F_2}(h(x))} \le 
		\sigma xg_2(x)\frac{p_2(h(x))}{\bar{F_2}(h(x))} \to 0, \quad \text{as } x\to +\infty.
	\end{equation}
	Therefore, $\lim_{x\to+\infty}\left(1-F_2(x)\right)/\left(1-F_1(f^{-1}(x))\right) = 1$.
\end{proof}


\end{document}